\documentclass[12pt]{iopart}
\newtheorem{prop}{Proposition}
\begin{document}

\title[ ]{Supersymmetric KdV equation: Darboux transformation and discrete systems}

\author{Ling-Ling Xue$^1$, D. Levi$^2$ and Q. P. Liu$^1$}

\address{$^1$ Department of Mathematics, China University of Mining and Technology, Beijing 100083, P.~R.~China}
\address{$^2$ Dipartimento di Matematica e Fisica,
Universita' degli Studi di Roma Tre and Sezione INFN Roma Tre,
Via della Vasca Navale  84,
I00146 Roma Italy}
\ead{qpl@cumtb.edu.cn}
\begin{abstract}
For the supersymmetric KdV equation, a proper Darboux transformation is presented. This Darboux transformation leads to the B\"{a}cklund transformation found early by Liu and Xie \cite{liu2}. The Darboux transformation and the related B\"{a}cklund transformation are used to construct integrable super differential-difference and difference-difference systems. The continuum limits of these discrete systems and of their Lax pairs are also considered.

\end{abstract}

\submitto{\JPA}
\maketitle

\section{Introduction}
In a seminal paper \cite{mr}, Manin and Radul proposed a supersymmetric KP hierarchy. This integrable hierarchy was later
studied intensively in \cite{mulase,uy}. As in the classical case, the reductions to a finite number of fields of  the supersymmetric KP hierarchy  have been studied. The simplest and  most important reduction is the supersymmetric KdV equation, which reads
\begin{equation}\label{skdv}
\alpha_t=\alpha_{xxx}+3(\alpha {\cal D}\alpha)_x,
\end{equation}
where $\alpha=\alpha(t,x,\theta)$ is a fermionic super field depending  on a temporal variable $t$ and the spatial super variables $(x,\theta)$ and ${\cal D}=\partial_\theta+\theta\partial_x$ is the corresponding super space derivative. $\theta$ is a fermionic commuting variable while  $x$ is a bosonic  anticommuting variable. Thus  the fermionic (odd) variable $\theta$ is nilpotent and we can carry out  the Taylor  expansion of the fermonic super field $\alpha$:
\[
\alpha=\rho(x,t)+\theta v(x,t),
\]
where the  component fields $\rho(x,t)$ and $v(x,t)$ are fermionic and bosonic quantities respectively. In this way, the supersymmetric KdV equation (\ref{skdv}) may be decomposed in the following system
\numparts
\begin{eqnarray}\label{skdvCa}
v_t=v_{xxx}+6vv_x-3\rho\rho_{xx},\\
\rho_t=\rho_{xxx}+3(v\rho)_x.\label{skdvCb}
\end{eqnarray}
\endnumparts
This system may be compared with the super KdV proposed by Kupershmidt \cite{BAK}
\numparts
\begin{eqnarray}\label{kuKDVa}
v_t=-v_{xxx}+6vv_x-3\rho\rho_{xx},\\
 \rho_t=-4\rho_{xxx}+6v\rho_x+3v_x\rho.\label{kuKDVb}
\end{eqnarray}\endnumparts
As observed by Mathieu \cite{ma}, even though the above two systems (2) and (3) appear  similar, they are very different. In fact, only (2) is invariant under
\[
\tilde v=\varepsilon\rho_x,\; \tilde\rho=\varepsilon v,
\]
where $\varepsilon$ is a fermionic parameter. Thus (2) is known as the supersymmetric KdV equation while (3) is the fermionic KdV equation.  More details on this may be found in \cite{ma} (see also \cite{BAK}).

   We can also write \eref{skdv} in potential form, i.e.
\begin{equation}\label{spkdv}
\beta_t=\beta_{xxx}+3\beta_x{\cal D}\beta_x,\quad\alpha=\beta_x.
\end{equation}
Like the KdV equation, the supersymmetric KdV equation has many interesting properties;  it is a bi-Hamiltonian system \cite{frm,op}, has a prolongation structure \cite{pro} and a B\"{a}cklund transformation \cite{liu2}, possesses an infinite number of conserved quantities and satisfies the Painlev\'{e} property \cite{ma,ma1}, can be studied in the framework of Hirota bilinear method \cite{bi2,lh,bi1}, etc. (see also \cite{fan,lou0,lou,fan0} for the most recent developments).

The existence of Darboux Transformations (DTs) is important  for integrable systems (see, for example, the monographs \cite{dl,gu,ms} and the recent review article \cite{cies}). In general, one can introduce two kinds of Darboux transformations, the elementary DT and the binary DT, and from them we get B\"{a}cklund transformations. For the supersymmetric KdV equation \eref{skdv}, the construction of DTs was initiated in \cite{liu95} and continued in \cite{lm98-2} and both elementary DT and binary DT have been constructed. Moreover, with the use  of a super Gardner transformation, one has derived a B\"{a}cklund transformation  for the supersymmetric KdV equation \cite{liu2}, which reads
\begin{eqnarray}\label{bt1}
&\left(\beta_{[1]}+\beta\right)_x-2 p_1 \left(\beta_{[1]}-\beta\right)+\frac{1}{2}(\beta_{[1]}-\beta){\mathcal D}\left(\beta_{[1]}-\beta\right)=0,
\end{eqnarray}
where $p_1 $ is the B\"{a}cklund parameter. Moreover, \eref{bt1} yields a superposition formula which may be applied to construct multi-soliton solutions.

Now it seems natural to ask ourselves which DT provides the  B\"{a}cklund transformation \eref{bt1}?
A careful consideration shows that none of DTs constructed in \cite{liu95,lm98-2} gives the B\"{a}cklund transformation \eref{bt1}. Thus, a proper DT is missing for the supersymmetric KdV equation and one of the purposes of this paper is to fill this gap.

Apart from their importance in the construction of solutions, Darboux and B\"{a}cklund transformations are known to play a key role for constructing nonlinear integrable discrete equations. Indeed, as showed by Levi and Benguria \cite{lb}, Darboux and B\"{a}cklund transformations can be used to provide integrable discretizations of the  continuous nonlinear integrable systems (see also \cite{le,schief,nqc,qncv}). Most recently, Grahovski and Mikhailov found integrable discretizations for a class of nonlinear Schr\"{o}dinger equations on Grassmann algebras \cite{sasha}.
A generalization of this approach to  supersymmetric integrable systems is up to now lacking and another purpose of this paper is to fill this gap. We will apply this idea to the supersymmetric KdV equation and show that the proper Darboux transformation  enables us to discretize it and to obtain an integrable  super differential-difference system and an integrable super difference-difference system.

The paper is organized as follows. In  Section 2 we present a proper Darboux transformation for the supersymmetric KdV equation and derive its well known B\"{a}cklund transformation. Then in Section 3 we use the obtained transformations to construct discrete integrable super systems. Both differential-difference equations and difference-difference equations are obtained. In Section 4, by performing various continuum limits, we show that our discrete systems are the proper discretizations of the potential supersymmetric KdV equation. The final Section summarizes the results and indicates the further directions of research. We leave to an Appendix  the proof that the obtained nonlinear partial difference equation satisfies the compatibility around the cube but does not have the tetrahedron property.

%
%

\section{Darboux transformations for the supersymmetric KdV}
Our result is summarized by the following Proposition.
\begin{prop}
Let us suppose that the super field $\phi(x,\lambda)$ satisfies the linear supersymmetric equation
\begin{eqnarray}\label{spx}
\phi_{xx}+ \alpha\mathcal{D}\phi=\lambda^2 \phi,
\end{eqnarray}
and  $\phi_{[0]}(x,p_1)$ be a solution of \eref{spx} for $\lambda=p_1$. Let us define
\begin{eqnarray}
\phi_{[1]}&\equiv&\left(\partial_x-\Lambda {\cal D}+p_1 \right)\phi,\label{phi1}\\
\alpha_{[1]}&\equiv& \alpha+2\Lambda_x,\;\;
\Lambda \equiv \frac{\phi_{[0],x}+p_1 \phi_{[0]}}{{\cal D}\phi_{[0]}}\label{field}.
\end{eqnarray}
Then ${\phi_{[1]}}(x,\lambda)$ will satisfy the following linear supersymmetric equation
\begin{eqnarray} \label{field1}
{\phi_{[1]}}_{xx}+ \alpha_{[1]} \left(\mathcal{D}\phi_{[1]}\right)=\lambda^2 \phi_{[1]} .
\end{eqnarray}
\end{prop}
{\it Proof}. It is just a trivial direct calculation in which (\ref{phi1}, \ref{field}) is introduced in \eref{field1} and $\phi$ it is shown to satisfy \eref{spx}.

The above DT (\ref{phi1}, \ref{field}) may be regarded as a nontrivial extension of the DT given in \cite{liu95}. Indeed, in Proposition 2.1 of \cite{liu95} one has considered  a solution of \eref{spx} with zero energy (namely $p_1 =0$); here our seed solution $\phi_{[0]}$ is an (odd) solution of \eref{spx} for an {\it arbitrary} value of $p_1 $.

We show now that the  DT  (\ref{phi1}, \ref{field})                                                                                                                                                                                             leads to the B\"{a}cklund transformation (\ref{bt1}) derived by other techniques in \cite{liu2}. In fact, let us
introduce the potential $\beta_{[1]}$ such that
$
\alpha_{[1]}={\beta_{[1]}}_x,
$
then  the first equation in \eref{field} gives
\begin{eqnarray}\label{Lambda1}
\Lambda=\frac{1}{2}\left(\beta_{[1]}-\beta\right),
\end{eqnarray}
where the constant of integration is set to zero.
On the other hand, taking into account that
\[
\phi_{[0],x}=\Lambda(\mathcal{D}\phi_{[0]})-p_1 \phi_{[0]}
\]
and
\[
\phi_{[0],xx}={p_1}^2\phi_{[0]}-\alpha(\mathcal{D}\phi_{[0]}),
\]
we obtain
\[
\Lambda_x+\beta_x-2p_1 \Lambda+\Lambda\left({\mathcal D}\Lambda\right)=0
\]
i.e. the B\"{a}cklund transformation \eref{bt1} for the supersymmetric KdV \eref{skdv}.

\section{Discrete equations from the super symmetric DT}
Let us rewrite the Darboux transformation (\ref{phi1}, \ref{field}) in matrix form. Introducing the vector $\Psi=(\phi,\phi_x,\mathcal{D}\phi,\mathcal{D}\phi_x)^{\texttt{T}}$,
with the help of \eref{bt1} one may rewrite \eref{spx} and \eref{phi1}  as the following systems
\begin{eqnarray}
\fl \Psi_x=L\Psi,\label{psix}\quad
 &L=
\left(
  \begin{array}{cccc}
    0 & 1 & 0 & 0 \\
    \lambda^2 & 0 & -\beta_x & 0 \\
    0 & 0 & 0 & 1 \\
    0 & \beta_x & \lambda^2-\mathcal{D}\beta_x & 0
  \end{array}
\right),\\
\fl \Psi_{[1]}=W\Psi,\label{psi1} \quad
&W=
\left(
  \begin{array}{cccc}
    p_1         & 1 & -\Lambda       & 0 \\
    \lambda^2 & p_1 &( \mathcal{D}\Lambda-2p_1)\Lambda  & -\Lambda\\
    0 & \Lambda & p_1-\mathcal{D}\Lambda & 1 \\
    \lambda^2\Lambda
    &(2p_1 -\mathcal{D}\Lambda)\Lambda
     & \lambda^2-2p_1(\mathcal{D}\Lambda)+(\mathcal{D}\Lambda)^2
     & p_1-\mathcal{D}\Lambda
  \end{array}
\right),
\end{eqnarray}
where $\Lambda$ is given by \eref{Lambda1}.
The compatibility of the two linear systems (\ref{psix}, \ref{psi1}) is
\begin{eqnarray} \label{eq1}
W_x+WL-L_{[1]}W=0,
\end{eqnarray}
 equivalent to  the B\"{a}cklund transformation \eref{bt1}. It is worthwhile to notice here that the matrix $W$ is a function, through $\Lambda$, of both $\beta$ and $\beta_{[1]}$. We can interpret \eref{eq1} as a differential-difference equation by the following identifications
\begin{eqnarray*}
\beta\equiv \beta_{n}(x),\quad \beta_{[1]} \equiv \beta_{n+1}(x).
\end{eqnarray*}
\eref{eq1} will admit as a Lax pair \eref{psix} and \eref{psi1}, therefore it will be integrable, at least in the Lax sense.

Let us introduce a new Darboux transformation
\begin{eqnarray}
\Psi_{[2]}=V\Psi,\label{psi2}
\end{eqnarray}
where the matrix $V$ is equal to the matrix $W$
with $p_1$ and $\beta_{[1]}$ replaced by $p_2$ and $\beta_{[2]}$ respectively, i.e. the matrix $V$ is a function through $\Lambda$ of both $\beta$ and $\beta_{[2]}$. If we consider the compatibility of \eref{psi1} with \eref{psi2}, i.e. the Bianchi permutability of a B\"acklund transformation \eref{eq1} of parameter $p_1$ with one of parameter $p_2$,
$(\Psi_{[1]})_{[2]}=(\Psi_{[2]})_{[1]}$, with $\beta_{[12]}=\beta_{[21]}$,
we get that the following  consistency condition
\begin{eqnarray} \label{eq2}
W_{[2]}V=V_{[1]}W
\end{eqnarray}
must be true. \eref{eq2} leads to
\begin{eqnarray}\label{superposition}
{\beta_{[12]}}=\beta+\frac{2(p_1+p_2)\left(\beta_{[1]}-\beta_{[2]}\right)}
{2(p_2-p_1)+\mathcal{D}\left(\beta_{[1]}-\beta_{[2]}\right)}
\end{eqnarray}
and
\begin{eqnarray}\label{condition}
\fl\left(\beta_{[1]}-\beta_{[2]}\right)\Big[2\left({\beta_{[1]}}-{\beta_{[2]}}\right)_x
+\left(\beta_{[1]}-\beta\right)\left(4 (p_2- p_1)+\mathcal{D}(\beta_{[1]}-\beta_{[2]})\right)\Big]=0.
\end{eqnarray}
\eref{superposition} is the superposition formula considered in \cite{liu2} while \eref{condition} is an extra constraint.

To obtain a partial difference  system, we use the following identifications
\begin{eqnarray*}
\beta\equiv \beta_{n,m}(x),\quad \beta_{[1]} \equiv \beta_{n+1,m}(x),\quad \beta_{[2]}\equiv \beta_{n,m+1}(x),\quad \beta_{[12]}\equiv \beta_{n+1,m+1}(x),
\end{eqnarray*}
thus the system (\ref{superposition}, \ref{condition}) can be interpreted as an integrable differential-partial difference system as it contains the superderivative $\mathcal D$.

To find a differential-difference system,  we rewrite the super fields in terms of its components. Let us assume that
\[\beta=\xi+\theta u, \quad \phi=\varphi+\theta f,
\]
and define
\begin{eqnarray} \nonumber
\xi\equiv \xi_{n}(x),\quad \;\xi_{[1]}\equiv \xi_{n+1}(x),\;\quad u\equiv u_{n}(x),\quad u_{[1]}\equiv u_{n+1}(x).
\end{eqnarray}
Then
the B\"{a}cklund transformation \eref{eq1} is split into
\numparts
\begin{eqnarray}\label{xi_hat}
(\xi_{n+1}+\xi_{n})_x=2p_1(\xi_{n+1}-\xi_{n})-\frac{1}{2}(u_{n+1}-u_{n})(\xi_{n+1}-\xi_{n}),\\
(u_{n+1}+u_{n})_x=2p_1(u_{n+1}-u_{n})-\frac{1}{2}(u_{n+1}-u_{n})^2-(\xi_{n+1}-\xi_{n})(\xi_{n})_x\label{u_hat},
\end{eqnarray}
\endnumparts
which is a differential-difference system.

Introducing $\chi_{n}=((\varphi_{n}),(\varphi_{n})_x,(f_{n}),(f_{n})_x)^\texttt{T}$, it follows from \eref{spx}, \eref{phi1} and (18) that
\begin{eqnarray}\label{LAXL}
\fl(\chi_{n})_x=\mathcal{L}_{n}\chi_{n},\quad
\mathcal{L}_{n}=
\left(
  \begin{array}{cccc}
    0 & 1 & 0 & 0 \\
    \lambda^2 & 0 & -(\xi_{n})_x & 0 \\
    0 & 0 & 0 & 1 \\
    0 & (\xi_{n})_x& \lambda^2-(u_{n})_x& 0 \\
  \end{array}
\right),
\end{eqnarray}
and
\begin{eqnarray}\label{lax1}
\fl \chi_{n+1}=\mathcal{W}_{n}\chi_{n},\quad
\mathcal{W}_{n}=
\left(
  \begin{array}{cccc}
    p_1         & 1 & -\eta       & 0 \\
    \lambda^2 & p_1 &-2{p_1} \eta+{\eta}g  & -{\eta} \\
    0 & \eta & p_1-g & 1 \\
    \lambda^2{\eta}
    &2{p_1} {\eta}-{\eta}g
     & \lambda^2-2{p_1}g+g^2
     & p_1-g\\
  \end{array}
\right)
\end{eqnarray}
with
\[
\eta\equiv\frac{1}{2}\left(\xi_{n+1}-\xi_{n}\right),\quad g\equiv\frac{1}{2}(u_{n+1}-u_{n}).
\]
Then the compatibility condition of \eref{LAXL} and \eref{lax1} gives
\begin{eqnarray*}
(\mathcal{W}_{n})_x+\mathcal{W}_{n}\mathcal{L}_{n}-\mathcal{L}_{n+1}\mathcal{W}_{n}=0
\end{eqnarray*}
which holds if and only if (18) is satisfied.
In the next Section, we will show that, after a continuum limit, the system (18) leads  to the potential supersymmetric KdV equation, thus it constitutes a discretization of the potential supersymmetric KdV equation.

To find a difference-difference system,  we  define
\begin{eqnarray*}
&\xi\equiv \xi_{n,m},\quad \;\xi_{[1]}\equiv \xi_{n+1,m},\;\quad \xi_{[2]}\equiv \xi_{n,m+1},\;\quad \xi_{[12]}\equiv \xi_{n+1,m+1},\\
&u\equiv u_{n,m},\quad u_{[1]}\equiv u_{n+1,m},\quad u_{[2]}\equiv u_{n,m+1},\quad u_{[12]}\equiv u_{n+1,m+1}.
\end{eqnarray*}

As above, we consider another Darboux transformation, equivalent to \eref{psi2}, which introducing $\chi_{n,m}=((\varphi_{n,m}),(\varphi_{n,m})_x,(f_{n,m}),(f_{n,m})_x)^\texttt{T}$, now reads
\begin{eqnarray}\label{lax2}
\chi_{n,m+1}=\mathcal{V}_{n,m}\chi_{n,m},
\end{eqnarray}
where the matrix $\mathcal{V}_{n,m}$ is the matrix $\mathcal{W}_{n}=\mathcal{W}_{n,m}$ of \eref{lax1}
with $p_1$,  $\xi_{n+1}=\xi_{n+1,m}$ and $u_{n+1}=u_{n+1,m}$ replaced by $p_2$,  $\xi_{n,m+1}$ and $u_{n,m+1}$ respectively.

Now the compatibility condition of \eref{lax1}, written as $\chi_{n+1,m}=\mathcal{W}_{n,m}\chi_{n,m}$,  and \eref{lax2}, namely
\begin{eqnarray*}
\mathcal{W}_{n,m+1}\mathcal{V}_{n,m}=\mathcal{V}_{n+1,m}\mathcal{W}_{n,m}
\end{eqnarray*}
yields an integrable difference-difference system
\numparts
\begin{eqnarray}
\fl\xi_{n+1,m+1}=\xi_{n,m}+\frac{2(p_1+p_2)\left(\xi_{n+1,m}-\xi_{n,m+1}\right)}
{2(p_2-p_1)+u_{n+1,m}-u_{n,m+1}},\label{xi12}\\
\fl u_{n+1,m+1}=u_{n,m}+\frac{2(p_1+p_2)(u_{n+1,m}-u_{n,m+1})}
{2(p_2-p_1)+u_{n+1,m}-u_{n,m+1}}\nonumber\\
\fl\quad -\frac{(p_1+p_2)(4(p_2-p_1)+u_{n+1,m}-u_{n,m+1})}
{\left(2(p_2-p_1)+u_{n+1,m}-u_{n,m+1}\right)^2}\left(\xi_{n+1,m}-\xi_{n,m+1}\right)\left(\xi_{n,m+1}-\xi_{n,m}\right).\label{u12}
\end{eqnarray}
\endnumparts
In the next Section, we will show that, after a double continuum limit, the system (22) leads to the potential supersymmetric KdV equation, thus it constitutes  a discretization of the potential supersymmetric KdV equation.

\noindent
{\bf Remarks}:
\begin{enumerate}
\item  The difference-difference system (22) may also be derived from \eref{superposition} and \eref{condition} directly. Indeed, one can write  \eref{superposition}  in components and obtain \eref{xi12}
and
\begin{eqnarray}\label{u120}
u_{n+1,m+1}&=&u_{n,m}+\frac{2(p_1+p_2)(u_{n+1,m}-u_{n,m+1})}
{2(p_2-p_1)+u_{n+1,m}-u_{n,m+1}}  \\ \nonumber
&+&\frac{2(p_1+p_2)\left(\xi_{n+1,m}-\xi_{n,m+1}\right)\left(\xi_{n+1,m}-\xi_{n,m+1}\right)_x}
{(2(p_2-p_1)+u_{n+1,m}-u_{n,m+1})^2}.
\end{eqnarray}
\eref{condition} gives
\begin{eqnarray*}
2\left(\xi_{n+1,m}-\xi_{n,m+1}\right)\left(\xi_{n+1,m}-\xi_{n,m+1}\right)_x\\
=-\left(\xi_{n+1,m}-\xi_{n,m+1}\right)\left(\xi_{n,m+1}-\xi_{n,m}\right)(4(p_2-p_1)+u_{n+1,m}-u_{n,m+1}),
\end{eqnarray*}
from which we can eliminate the differential part of \eref{u120} and obtain \eref{u12}.

\item If $\xi=0$, (22) reduces to the well-known lattice potential KdV equation
\begin{eqnarray*}
&u_{n+1,m+1}=u_{n,m}+\frac{2(p_1+p_2)(u_{n+1,m}-u_{n,m+1})}
{2(p_2-p_1)+u_{n+1,m}-u_{n,m+1}}.
\end{eqnarray*}

\item The system (22) is integrable in the sense that it possesses Lax representation. In the appendix, it will be shown that this system is consistent around a cube, i.e. it has the CAC property \cite{abs1}.
\end{enumerate}

\section{Continuum limits of the discrete equations obtained in the previous Section}
We can characterize the discrete systems obtained in the previous Section by analyzing their continuum limits \cite{nc,frank}. We obtained the differential-difference system (18) and the difference-difference system (22) as discrete versions of the potential supersymmetric KdV equation. We now justify this claim by considering their continuum limits.
\subsection{The continuous limit of (18)}\label{xiuhatp}
To carry out the continuous limit of (18) we define the new continuous variable $\tau$ as
\begin{eqnarray*}
\xi_{n}(x)\equiv \xi(x,\tau),\quad u_{n}(x)\equiv u(x,\tau),\quad \tau=\frac{n}{p_1}.
\end{eqnarray*}
Then
\begin{eqnarray*}
\xi_{n+1}(x)\equiv \xi\left(x,\tau+\frac{1}{p_1}\right),\quad u_{n+1}(x)\equiv u\left(x,\tau+\frac{1}{p_1}\right),
\end{eqnarray*}
can be expanded in $\frac{1}{p_1}$, and defining a new independent variable  $t$ in term of $\tau$ and $x$ such that
\begin{eqnarray*}
\partial_\tau=\partial_x+\frac{1}{12{p_1}^2}\partial_t,
\end{eqnarray*}
 we obtain in the continuous limit up to terms of order $\frac{1}{p_1}$
\begin{eqnarray}\label{spkdv_c}
\xi_t=\xi_{xxx}+3\xi_xu_x,\quad u_t=u_{xxx}+3u_{x}^2+3\xi_{xx}\xi_{x}
\end{eqnarray}
which is the potential form of supersymmetric KdV equation \eref{spkdv}.

\subsection{The semi-continuous limits of  (22)}
We present here two different results obtained by implementing different continuous limits of (22), at first when we send to infinity just one  of the discrete variables and secondly when we send to infinity a combination of both discrete variables.
\subsubsection{Straight continuum limit}
The system (22)  may be regarded as a discrete analogue
of the differential-difference system (18). To see it, let us
define
\begin{eqnarray*}
\xi_{n,m}\equiv \xi_{n}(x),\quad u_{n,m}\equiv u_{n}(x),\quad x=\frac{m}{p_2}.
\end{eqnarray*}
For $\frac{1}{p_2}$ small, we have the following Taylor series expansions
\begin{eqnarray*}
&\xi_{k,m+1}=\xi_{k}\left(x+\frac{1}{p_2}\right)=\xi_{k}+\frac{1}{p_2}\xi_{k,x}+O\left(\frac{1}{{p_2}^2}\right),\\
&u_{k,m+1}=u_{k}\left(x+\frac{1}{{p_2}}\right)=u_{k}+\frac{1}{{p_2}}u_{k,x}+O\left(\frac{1}{{p_2}^2}\right),
\end{eqnarray*}
where $k$ can take the value $n$ or $n+1$.
Substituting the above expansions into (22), the leading terms yield
\numparts
\begin{eqnarray}
&(\xi_{n+1}+\xi_n)_x=2p_1(\xi_{n+1}-\xi_n)-\frac{1}{2}(\xi_{n+1}-\xi_n)(u_{n+1}-u_n),\\
&(u_{n+1}+u_n)_x=2p_1(u_{n+1}-u_n)-\frac{1}{2}(u_{n+1}-u_n)^2-(\xi_{n+1}-\xi_n)\xi_{n,x},
\end{eqnarray}
\endnumparts
i.e. (18).

\subsubsection{Skew continuum limit}

In this case, we introduce the new discrete variable
\begin{eqnarray} \label{skew}
N=n+m,\quad p_2=p_1+\epsilon ,\quad \tau=\epsilon m,
\end{eqnarray}
and
\begin{eqnarray} \label{skew1}
\xi_{n,m} &\equiv \xi_{N}(\tau),\qquad \qquad\qquad u_{n,m}&\equiv u_{N}(\tau),\\ \nonumber
\xi_{n+1,m} &\equiv \xi_{N+1}(\tau),\qquad \qquad u_{n+1,m}&\equiv u_{N+1}(\tau),\\ \nonumber
\xi_{n,m+1} &\equiv \xi_{N+1}(\tau+\epsilon),\qquad\; u_{n,m+1}&\equiv u_{N+1}(\tau+\epsilon),\\ \nonumber
\xi_{n+1,m+1} &\equiv \xi_{N+2}(\tau+\epsilon),\;\quad u_{n+1,m+1}&\equiv u_{N+2}(\tau+\epsilon),
\end{eqnarray}
and develop all dependent variables in Taylor series in $\epsilon $. After inserting them into (22), the leading order terms give
\begin{eqnarray*}
\fl\xi_{N+1,\tau}=-\frac{1}{4p_1}(\xi_{N+2}-\xi_{N})(2-u_{N+1,\tau}),\\
\fl(u_{N+2}-u_{N})(2-u_{N+1,\tau})^2+4p_1u_{N+1,\tau}(2-u_{N+1,\tau})=2p_1\xi_{N+1,\tau}(4-u_{N+1,\tau})(\xi_{N+1}-\xi_{N}).
\end{eqnarray*}
Solving this system with respect to $\xi_{N+1,\tau}$ and $u_{N+1,\tau}$, and shifting $N+1$ to $N$, we get, either
\numparts
\begin{eqnarray}
\fl\xi_{N,\tau}=\frac{2(\xi_{N+1}-\xi_{N-1})}{u_{N+1}-u_{N-1}-4p_1}, \label{tau1a}\\
\fl u_{N,\tau}=\frac{2(u_{N+1}-u_{N-1})}{u_{N+1}-u_{N-1}-4p_1}
+\frac{(u_{N+1}-u_{N-1}-8p_1)}{(u_{N+1}-u_{N-1}-4p_1)^2}(\xi_{N+1}-\xi_{N-1})(\xi_{N}-\xi_{N-1}), \label{tau1b}
\end{eqnarray}
\endnumparts
or the trivial solution
\begin{eqnarray*}
\xi_{N,\tau}=0, \quad u_{N,\tau}=2.
\end{eqnarray*}

If the fermionic variable $\xi_{N}$ is null, the system (28) is reduced to
\begin{eqnarray*}
&u_{N,\tau}=\frac{2(u_{N+1}-u_{N-1})}{u_{N+1}-u_{N-1}-4p_1},
\end{eqnarray*}
a differential-difference equation related to the Kac-van Moerbeke
equation \cite{kvm}. Thus, with this limit, the system (28) is a super extension of the Kac-van Moerbeke equation.

\subsection{Full continuum limit}
Two different semi-continuum limits have been considered for  (22) and two
  differential-difference systems, namely (25) and (28), are obtained. As (25) is equal to (18) its continuous limit is contained in Section \eref{xiuhatp} and it leads to the potential supersymmetric KdV equation \eref{spkdv}. In the following we will consider the continuum limit of (28) and  it will turn out that it also leads to the potential  supersymmetric KdV equation \eref{spkdv}.

For (28),
defining
\begin{eqnarray*}
\xi_{N}(\tau)\equiv \xi(s,\tau),\quad u_{N}(\tau)\equiv u(s,\tau),\quad s=\frac{N}{p_1},
\end{eqnarray*}
expanding
\begin{eqnarray*}
\xi_{N\pm1}(\tau)\equiv \xi\left(s\pm\frac{1}{p_1},\tau\right),\quad u_{N\pm1}(\tau)\equiv u\left(s\pm\frac{1}{p_1},\tau\right)
\end{eqnarray*}
in $\frac{1}{p_1}$, and redefining the independent variables from $s,\,\tau$ to $x,\,t$
\begin{eqnarray*}
&\partial_s=\partial_x,\quad\partial_\tau=-\frac{1}{{p_1}^2}\partial_x-\frac{1}{6{p_1}^4}\partial_t,
\end{eqnarray*}
we get once again \eref{spkdv_c}.

\subsection{Semi-continuous limits of the Lax pair }

The continuum limits considered above were done on the level of the nonlinear systems. We actually can work out also the continuum limits of the corresponding  Lax pair \eref{lax1} and \eref{lax2}:
\begin{eqnarray}\label{lax}
(p_1-\lambda)\chi_{n+1,m}=\mathcal{W}_{n,m}\chi_{n,m},\quad
(p_2-\lambda)\chi_{n,m+1}=\mathcal{V}_{n,m}\chi_{n,m},\quad
\end{eqnarray}
in which the matrices $\mathcal{W}$ and $\mathcal{V}$ are defined in Section 3.
We now perform at first the skew continuum limit \eref{skew} on the Lax pair in order to obtain a Lax pair for (28).
Using (\ref{skew}, \ref{skew1}) the first element of the Lax pair \eref{lax} reads
\begin{eqnarray}\label{chiN+1}
\fl (p_1-\lambda)\chi_{N+1}=\mathcal{W}_N\,\chi_N,\quad
\mathcal{W}_{_N}=
\left(
  \begin{array}{cccc}
    p_1         & 1 & -\zeta       & 0 \\
    \lambda^2 & p_1 &(  h-2p_1 )\zeta  & -\zeta  \\
    0 & \zeta  & p_1-h  & 1 \\
    \lambda^2\zeta
    &(2p_1   -h)\zeta
     & \lambda^2+(  h-2p_1 )h
     & p_1-h
  \end{array}
\right),
\end{eqnarray}
where
$
\zeta \equiv \frac{1}{2}(\xi_{N+1}-\xi_N),\, h\equiv \frac{1}{2}(u_{N+1}-u_N).
$
Expanding the second element of the Lax pair \eref{lax} in $\epsilon$, we obtain
\begin{eqnarray*}
(\epsilon+p_1-\lambda)(\chi_{N+1}+\epsilon\chi_{N+1,\tau}+\cdots)
=\mathcal{W}_N \,\chi_N
+\epsilon\mathcal{T}_{N}\chi_N
+\cdots,
\end{eqnarray*}
where
\begin{eqnarray*}
&\mathcal{T}_N=\left(
  \begin{array}{cccc}
    1         & 0 & -\frac{1}{2}\xi_{N+1,\tau}       & 0 \\
    0 & 1 & t_{23} & -\frac{1}{2}\xi_{N+1,\tau} \\
    0 & \frac{1}{2}\xi_{N+1,\tau}  & 1-\frac{1}{2}u_{N+1,\tau} & 0 \\
    \frac{1}{2}\lambda^2\xi_{N+1,\tau}
    &-t_{23}
     &t_{43}
     & 1-\frac{1}{2}u_{N+1,\tau}\\
  \end{array}
\right)
\end{eqnarray*}
with
\begin{eqnarray*}
&t_{23} \equiv -p_1 \xi_{N+1,\tau}-2\zeta+\frac{1}{2}\left(h\xi_{N+1,\tau}+\zeta u_{N+1,\tau}\right),\\
&t_{43} \equiv -p_1 u_{N+1,\tau}-2h+hu_{N+1,\tau}.
\end{eqnarray*}
Next with the help of \eref{chiN+1} we obtain as the coefficient of the leading term of order
$O(\epsilon)$ the following equation
\begin{eqnarray}\label{chitau}
(p_1-\lambda)\chi_{N+1,\tau}=\mathcal{T}_N\chi_N-\chi_{N+1}.
\end{eqnarray}
The consistency condition of  \eref{chiN+1} and \eref{chitau}
\begin{eqnarray*}
\mathcal{W}_{N,\tau}\mathcal{W}_{N-1}+\mathcal{W}_N\mathcal{T}_{N-1}-\mathcal T_N\mathcal{W}_{N-1}=0
\end{eqnarray*}
leads to (28),
thus (\ref{chiN+1}, \ref{chitau}) constitute a Lax pair for the supersymmetric Kac-van Moerbeke
equation (28).

\section{Conclusion}
In this paper, we give a new Darboux transformation for the supersymmetric KdV equation. By means of this Darboux transformation, the supersymmetric KdV equation is discretized. Both differential-difference system and difference-difference system are obtained, their integrability is shown and various continuum limits are considered.

This work opens the way to the construction of many discrete supersymmetric equations. We are planning to apply the idea to other supersymmetric integrable equations. In particular we are considering the supersymmetric Schr\"{o}dinger equation \cite{snls} and the Kupershimidt's super KdV equation \cite{BAK}, which is  a fermionic rather than supersymmetric extension of the KdV equation.

\section*{Acknowledgments}

LLX and QPL   are supported by the National
Natural Science Foundation of China (grant numbers: 10971222, 11271366 and 11331008) and the Fundamental Research Funds for Central Universities. LD  has been partly supported by the Italian Ministry of Education and Research, PRIN ``Continuous and discrete nonlinear integrable evolutions: from water
waves to symplectic maps" and  thanks the  China University of Mining and Technology for its support during his visit to Beijing where this work has been carried out.

\section*{Appendix: CAC property of the system (22)}
An integrable partial difference equation defined on a square lattice is  characterized by the Consistency-Around-the-Cube (CAC) property \cite{abs1,hietar1,hietar2}, equivalent to the Bianchi permutability theorem for B\"acklund transformations of integrable nonlinear partial differential equations.  In the following we will show that also the system (22) possesses such a property.

From (22), by adding a third direction, we have the following equations
\begin{eqnarray*}
\fl \xi_{[ij]}\;=\xi+f_{[ij]}\left(\xi_{[i]}-\xi_{[j]}\right),\quad
u_{[ij]}=u+f_{[ij]}\left(u_{[i]}-u_{[j]}\right)+g_{[ij]}\left(\xi_{[i]}-\xi_{[j]}\right)\left(\xi_{[j]}-\xi\right),
\end{eqnarray*}
where
\begin{eqnarray*}
&f_{[ij]}\equiv\frac{2(p_i+p_j)}{2(p_j-p_i)+u_{[i]}-u_{[j]}},\quad
g_{[ij]}\equiv-\frac{(p_i+p_j)\left(4(p_j-p_i)+u_{[i]}-u_{[j]}\right)}
{\left(2(p_j-p_i)+u_{[i]}-u_{[j]}\right)^2}
\end{eqnarray*}
with $1\leq i,j\leq 3$.
Then we obtain
\begin{eqnarray*}
\fl\xi_{[ijl]}=\xi_{[l]}+f_{[ijl]}\left(\xi_{[il]}-\xi_{[jl]}\right),\quad
u_{[ijl]}=u_{[l]}+f_{[ijl]}\left(u_{[il]}-u_{[jl]}\right)+g_{[ijl]}\left(\xi_{[il]}-\xi_{[jl]}\right)\left(\xi_{[jl]}-\xi_{[l]}\right),
\end{eqnarray*}
where
\begin{eqnarray*}
\fl f_{[ijl]}\equiv\frac{2(p_i+p_j)}{2(p_j-p_i)+u_{[il]}-u_{[jl]}},\quad
g_{[ijl]}\equiv-\frac{(p_i+p_j)\left(4(p_j-p_i)+u_{[il]}-u_{[jl]}\right)}
{\left(2(p_j-p_i)+u_{[il]}-u_{[jl]}\right)^2}.
\end{eqnarray*}
Consistency means that $\xi_{[123]}=\xi_{[231]}=\xi_{[312]},\,u_{[123]}=u_{[231]}=u_{[312]}$.
Indeed, by direct calculation we find
\begin{eqnarray*}
\fl\xi_{[123]}&=\frac{1}{H}\sigma_{ijl}\xi_{[i]}(p_i+p_j)(p_i+p_l)(2(p_j-p_l)-u_{[j]}+u_{[l]})\\
\fl &+\frac{1}{2H^2}\left(\xi_{[1]}-\xi_{[3]}\right)\left(\xi_{[2]}-\xi_{[3]}\right)
\left(\xi_{[3]}-\xi\right)(p_1+p_2)(p_1+p_3)(p_2+p_3)\sigma_{ijl}(p_i-p_j)u_{[l]},\\
\fl u_{[123]}& =\frac{1}{H}\sigma_{ijl}p_{i}^2\left[u_{[i]}\left(u_{[l]}-u_{[j]}\right)+2p_j\left(u_{[i]}-u_{[j]}+u_{[l]}\right)-2p_l\left(u_{[i]}+u_{[j]}-u_{[l]}\right)\right]\\
\fl& +\frac{1}{2H^2}(p_1+p_2)(p_1+p_3)(p_2+p_3)\\
\fl&\times\left\{
\sigma_{ijl}\xi\xi_{[i]}\left(2(p_l-p_j)+u_{[j]}-u_{[l]}\right)
\left[p_i\left(u_{[j]}-u_{[l]}\right)+p_j\left(u_{[l]}-u_{[i]}\right)+p_l\left(u_{[i]}-u_{[j]}\right)\right]\right.\\
\fl& \left.+\sigma_{ijl}\xi_{[j]}\xi_{[l]}\left[8(p_i-p_j)(p_i-p_l)(p_j-p_l)+(p_j-p_l)\left(u_{[i]}-u_{[j]}\right)\left(u_{[i]}-u_{[l]}\right)\right.\right.\\
\fl&  \left.\left.-4u_{[i]}(p_j-p_l)(p_i-p_j-p_l)-2u_{[j]}(p_i-p_l)(p_i-p_j+2p_l)+2u_{[l]}(p_i-p_j)(p_i+2p_j-p_l)
\right]\right\},
\end{eqnarray*}
where \[H\equiv 2(p_1-p_2)(p_1-p_3)(p_2-p_3)+\sigma_{ijl}(p_{i}^2-p_{j}^2)u_{[l]},\]
and $\sigma_{ijl}$ denotes the cyclic sum over the subscripts $(i,\,j,\,l)=(1,2,3),(2,3,1),(3,1,2)$.
Then $\xi_{[123]}$ and $u_{[123]}$ are symmetric under any permutation of (1,\,2,\,3),
thus (22) obeys the CAC property. We notice that since both $\xi_{[123]}$ and $u_{[123]}$ are independent of $u$ but dependent on $\xi$, the system (22) does not have the tetrahedron property.

\end{document}